\documentclass[11pt]{article}

\def\mytitle#1{\setcounter{equation}{0}
\setcounter{footnote}{0}
\begin{flushleft}\Large\textbf{#1}\end{flushleft}
\vspace{0.25cm}}
\def\myname#1{\leftline{{\large #1}}\vspace{-0.13cm}}
\def\myplace#1#2{\small\begin{flushleft}\textit{#1}\\
\texttt{#2}\end{flushleft}}

\def\myclassification#1{\small\noindent
Pacs no :
#1\vspace{0.5cm}}
\usepackage{graphicx}% Include figure files
\begin{document}

\mytitle{Observational data fitting to constrain Variable Modified
Chaplygin Gas in the background of Horava-Lifshitz Gravity}

\vskip0.2cm \myname{Chayan
Ranjit\footnote{chayanranjit@gmail.com}} \myplace{Department of
Mathematics, Egra S. S. B. College, Purba Medinipur-721429, W.B.
India.}{} \vskip0.1cm

\vskip0.2cm \myname{Prabir Rudra\footnote{prudra.math@gmail.com}}
\myplace{Department of Mathematics, Asutosh College, Kolkata-700
026, India.}{}

\begin{abstract}
FRW universe in Horava-Lifshitz (HL) gravity model filled with a
combination of dark matter and dark energy in the form of variable
modified Chaplygin gas (VMCG) is considered. The permitted values
of the VMCG parameters are determined by the recent astrophysical
and cosmological observational data. Here we present the Hubble
parameter in terms of the observable parameters $\Omega_{dm0}$,
$\Omega_{vmcg0}$, $H_{0}$, redshift $z$ and other parameters like
$\alpha$, $A$, $\gamma$ and $n$. From Stern data set (12 points),
we have obtained the bounds of the arbitrary parameters by
minimizing the $\chi^{2}$ test. The best-fit values of the
parameters are obtained by 66\%, 90\% and 99\% confidence levels.
Next due to joint analysis with BAO and CMB observations, we have
also obtained the bounds of the parameters ($A,\gamma$) by fixing
some other parameters $\alpha$ and $n$. The best fit value of
distance modulus $\mu(z)$ is obtained for the VMCG model in HL
gravity, and it is concluded that our model is perfectly
consistent with the union2 sample data.
\end{abstract}

\myclassification{04.60.Pp, 98.80.Qc, 98.80.-k}

%%%%%%%%%%%%%%%%%%%%%%%%%%%%%%%%%%%%%%%%%%%%%%%%%%%%%%%%%%%%%%%%%%%%%%%%%%%%%%%
\section{Introduction}
%%%%%%%%%%%%%%%%%%%%%%%%%%%%%%%%%%%%%%%%%%%%%%%%%%%%%%%%%%%%%%%%%%%%%%%%%%%%%%%%
For several decades we have tried to give birth to a well
constructed quantum gravity theory that can reconcile with the
general theory of Relativity. With the discovery of the late
cosmic acceleration \cite{Perlmutter,Riess,Riess1,Bennet,Sperge}
at the turn of the last century, the unified theory became all the
more necessary. Several footsteps can be found in literature that
aims at producing a UV complete theory. Motivated by the success
of the theory proposed by Lifshitz \cite{Lifshitz} in solid state
physics, Horava proposed a gravity theory widely known the
Horava-lifshitz (HL) gravity \cite{Horava1}. Taking the UV limit
into account, HL gravity has a Lifshitz like anisotropic scaling
as $t\rightarrow l^{z} t$ and $x^i \rightarrow l x^i$ between
space and time. As this is characterized by the dynamical critical
exponent $z=3$, it breaks the Lorentz invariance, while in the
infra red limit, the scale reduces to $z=1$, i.e., it is reduced
to classical general relativistic theory of gravity in the low
energy limit. Even if we add a $\frac{1}{a^4}$ term, `$a$' being
the scale factor, with the Friedmann equation \cite{Lu, Cai,
Calcagni1, Calcagni2, Kiritsis1, Jamil1, Setare}  we will have HL
gravity equations.

In the original proposal, Horava \cite{Horava1, Horava2}
considered projectability condition with/without the detailed
balance. The detailed balance condition was originally proposed in
order to reduce the number of operations in the action (i.e.,
number of independent coupling constants) and to simplify some
properties of the quantum system. As a result the form of
potential in the 4-dimensional Lorentzian action is restricted to
a specific form in terms of a 3D Euclidean theory. But from
cosmological point of view, this condition leads to major
obstacles and hence should be abandoned.

On the other hand, the fundamental symmetry of the theory namely
the foliation-preserving diffeomorphism invariance leads to the
projectability condition. In particular, one should have 3D
spatial diffeomorphism  and the space-independent time
reparametrization. As the lapse function is essentially the Gauge
degree of freedom associated with the time reparametrization, so
it should be independent of space coordinates. This is termed as
projectibility condition. Due to this projectibility condition,
the Hamiltonian constraint is not a local equation satisfied at
each spatial point but an equation integrable over a whole space,
i.e., a global Hamiltonian constraint. Note that absence of
projectability condition and imposing a local version of the
Hamiltonian constraint lead to phenomenological obstacles and
theoretical inconsistencies.

However, the important drawback of projectibility condition is
that : due to the global Hamiltonian constraint the general
relativity could not be recovered from the Horava gravity for
arbitrary small lambda. Also the projectibility condition from the
point of view of condensed matter physics may not be appropriate
for describing the (quantum) gravity.

In extragalactic astronomy, type Ia supernovae have a
characteristic light curve. The similarity in absolute luminosity
profiles of nearly all known type Ia supernovae helps
astrophysicists to treat them as a standard candle. Using this
fact the observations Riess et al and Perlmutter et al in 1998
\cite{Perlmutter, Riess} led us to the conclusion that our
universe is going through a cosmic acceleration. There are two
popular ways that provide theoretical support to the present day
accelerated expansion of our universe. One is to modify the
geometric part of Einstein equation, leading to the ideas of
modified gravity and the other is to consider the universe to be
filled up uniformly by some exotic matter possessing negative
pressure. The Friedmann equation $\frac{\ddot{a}}{a}=-4\pi G
\left(\rho+3p\right)/3$ requires the condition
$\left(\rho+3p\right)<0$ for accelerated expansion ($\ddot{a}>0$).
As density is an ever positive physical quantity, we see the
Equation of state (EoS) parameter must be negative and also less
than $-1/3$. Such a negative pressure creating substance is aptly
termed as dark energy (DE here after). Recent observational data
indicates that DE occupies $73\%$ of the whole matter-energy of
our universe. Theoretically we can find many proposed DE
candidates. Variable modified Chaplygin gas (VMCG) is one among
them. The equation of state of VMCG is given as
\cite{Jamil,Farooq}
\begin{equation}\label{EoS}
p=\alpha\rho-\frac{\beta(a)}{\rho^{n}}
\end{equation}
Here we consider $\beta(a)=\beta_{0} a^{-\gamma}$, where
$\beta_{0}$ and $\gamma$ are constants.

In 1904, Chaplygin \cite{Chaplygin1} introduced Chaplygin gas
(CG), whose EoS is given by
\begin{equation}\label{chaplyginequation1}
p=-\frac{\beta}{\rho}
\end{equation}
where $\beta$ is a positive constant. An attractive feature of the
model is that it behaves as dust-like matter at an early stage and
as cosmological constant at later stages i.e. the CG behaves like
a pressure less fluid for small values of scale factor and as a
cosmological constant for large values of scale factors which tend
to accelerate the expansion. At the beginning of this century, the
CG model went through a series of modifications, all in the quest
of a suitable candidate for DE.

In \cite{Bento1}, Bento et al for the first time gave the idea of
Generalised CG having the EoS
\begin{equation}\label{chaplyginequation3}
p=-\frac{\beta}{\rho^{n}}
\end{equation}
Whereas in 2010, Jamil \cite{Jamil4} presented a model in which
the new generalized Chaplygin gas interacts with matter. Benaoum
in 2002 further modified CG model and proposed the Modified CG
(MCG) model obeying the equation of state
\cite{Benaoum1,Jamil2,Debnath,Jamil5}
\begin{equation}\label{chaplyginequation4}
p=\alpha \rho-\frac{\beta}{\rho^{n}}
\end{equation}
with $\alpha> 0$ and $0\leq n \leq 1$. When $\alpha=\frac{1}{3}$,
this EoS shows radiation era at one extreme (when the scale factor
is vanishingly small) while $\Lambda$CDM model at the other
extreme (when the scale factor is infinitely large). At all stages
it shows a mixture. Amidst these there also exist one stage when
the pressure vanishes and the matter content is equivalent to pure
dust. On further modification VMCG came into existence. In
\cite{Jamil3}, Jamil investigated the evolution of a Schwarzschild
black hole in the standard model of cosmology using the
phantom-like modified variable Chaplygin gas and the viscous
generalized Chaplygin gas.

%%%%%%%%%%%%%%%%%%%%%%%%%%%%%%%%%%Motive%%%%%%%%%%%%%%%%%%%%%%%%%%%%%%%%%%%%%%%%%%%%%%%%

Present day trend of literature says that the combination of DE
with modified gravity together gives more interesting results. In
2010 Park showed in \cite{Park} that the Friedmann equation in
Horava gravity contains additional $a^{-4}$, $a^{-2}$ and
cosmological constant terms as the effective DE and he predicted
that these terms may be responsible for cosmic acceleration. These
terms, coined as Horava effective DE, were tried to be constrained
on a few occasions \cite{Dutta, Ali}. The best ever approach was
via phenomenological parametrization of the relevant physical
quantities that have been well-studied. Park himself considered
the widely used CPL parametrization. B. C. Paul and his coleagues
have studied constraints for different exotic fluid model in the
back ground of HL gravity \cite{Paul1, Paul2}. In this work, we
shall study the limits of the DE parameters constrained by
different data sets. First we will set the constraint for the
closed universe and analyze it. Later on we deal with the range of
parameters for open and flat universe. It is quite understood that
any observational constraints on HL gravity do not enlighten the
discussion about the well-known conceptual problems and
instabilities of the theory, nor it can address the questions
concerning its validity. Therefore in the present analysis the HL
gravity has to be considered as a phenomenological model. The same
holds for the Chaplygin Gas model as well.

The paper is organized as follows: The basic generalized equations
for HL gravity are given in section (2). Various dimensionless
density parameters have been discussed in section (3). The main
mechanisms which will be followed to analyze the data is briefly
given in section (4). Lastly, a brief summary and a fruitful
conclusion have been drawn in section (5).

%%%%%%%%%%%%%%%%%%%%%%%%%%%%%%%%%%%%%%%%%%%%%%%%%%%%%%%%%%%%%%%%%%%%%%%%%%%%%%%%%%%%%55
\section{Basic Equations in Horava-Lifshitz Gravity}\label{Basic Calculations}
%%%%%%%%%%%%%%%%%%%%%%%%%%%%%%%%%%%%%%%%%%%%%%%%%%%%%%%%%%%%%%%%%%%%%%%%%%%%%%%%%%%%5

It is convenient to use the Arnowitt-Deser-Misner decomposition of
the metric which is given by \cite{Calcagni1, Calcagni2,
Kiritsis1}
\begin{equation}\label{HL1}
ds^2=-N^2dt^2+g_{i j} \left(dx^i+N^i dt\right)\left(dx^j+N^j dt\right).
\end{equation}
Here, $N$ is the lapse function, $N_i$ is the shift vector,
$g_{ij}$ is the spatial metric. The scaling transformation of the
coordinates reads as : $t\rightarrow l^3 t$ and $x^i\rightarrow l
x^i$. The HL gravity action has two constituents, namely, the
kinetic and the potential term as
\begin{equation}
S_g=S_k+S_v=\int dt d^3 x \sqrt{g} N\left(L_k + L_v\right)
\end{equation}
where, the kinetic term is given by
\begin{equation}
S_k=\int dt d^3 x \sqrt{g} N \left[\frac{2\left(K_{ij}K^{ij}
-\lambda K^2\right)}{\kappa^2}\right]
\end{equation}
where, the extrinsic curvature is given as
\begin{equation}
K_{ij}=\frac{\dot{g}_{ij}-\Delta_i N_j-\Delta_j N_i}{2N}
\end{equation}
The number of invariants, while working with the Lagrangian,
$L_v$, can be reduced due to its symmetric property
\cite{Horava2}. This symmetry actually is known as detailed
balance. Considering this detailed balance the expanded form of
the action becomes
\begin{equation}
S_g= \int dt d^3x \sqrt{g} N \left[\frac{2\left(K_{ij}K^{ij}
-\lambda K^2\right)}{\kappa^2}+\frac{\kappa^2
C_{ij}C^{ij}}{2\omega^4} -\frac{\kappa^2 \mu \epsilon^{i j k }
R_{i, j} \Delta_j R^l_k}{2\omega^2
\sqrt{g}}\right.$$$$\left.+\frac{\kappa^2 \mu^2 R_{ij} R^{ij}}{8}
-\frac{\kappa^2
\mu^2}{8(3\lambda-1)}\left\{\frac{(1-4\lambda)R^2}{4} +\Lambda R
-3 \Lambda^2 \right\}\right]
\end{equation}
here $C^{ij}=\frac{\epsilon^{ijk} \Delta_k\left(R_i^j-\frac{R}{4}
\delta^j_i\right)}{\sqrt{g}}$ is the Cotton tensor and all the
covariant derivatives are determined with respect to the spatial
metric. $g_{ij} \epsilon^{ijk}$ is a totally antisymmetric unit
tensor, $\lambda$ is a dimensionless constant and $\kappa$,
$\omega$ and $\mu$ are constants.

Now to set the matter-tensor the total energy density and
pressure, $\rho_m$ and $p_m$ respectively will be taken which will
contain in itself the impact of the baryonic density ($\rho_b$),
dark matter density ($\rho_{dm}$), etc.

Assuming only temporal dependency of the lapse function (i.e.,
$N\equiv N(t)$), Horava obtained a gravitational action. Using FRW
metric with $N=1~,~g_{ij}=a^2(t)\gamma_{ij}~,~N^i=0$ and
$$\gamma_{ij}dx^i dx^j=\frac{dr^2}{1-Kr^2}+r^2 d\Omega_2^2$$
where $K=-1,~1, 0$ represent open, closed and flat universe
respectively and taking variation of $N$ and $g_{ij}$ we obtain
the Friedmann equations \cite{Jamil1, Paul2}
\begin{equation}\label{HLFriedmann1}
H^2=\frac{\kappa^2}{6\left(3\lambda-1\right)}\left(\rho_m +\rho_r\right)
+\frac{\kappa^2}{6\left(3\lambda-1\right)}\left[\frac{3\kappa^2\mu^2 K^2}
{8\left(3\lambda-1\right)a^4}+\frac{3\kappa^2\mu^2 \Lambda^2}
{8\left(3\lambda-1\right)}\right]-\frac{\kappa^4 \mu^2 \Lambda K}{8\left(3\lambda-1\right)^2a^2}
\end{equation}
\begin{equation}\label{HLFriedmann2}
\dot{H}+\frac{3H^2}{2}=-\frac{\kappa^2}{4\left(3\lambda-1\right)}
\left(\rho_m w_m+\rho_r w_r\right)-\frac{\kappa^2}
{4\left(3\lambda-1\right)}\left[\frac{3\kappa^2\mu^2 K^2}
{8\left(3\lambda-1\right)a^4}+\frac{3\kappa^2\mu^2 \Lambda^2}
{8\left(3\lambda-1\right)}\right]-\frac{\kappa^4 \mu^2 \Lambda
K}{8\left(3\lambda-1\right)^2a^2}
\end{equation}
The term proportional to $\frac{1}{a^4}$ is an unique contribution
of HL gravity, which can be treated as ``Dark radiation term"
\cite{Calcagni1, Calcagni2, Kiritsis1} and the constant term is
the cosmological constant.

The conservation equation of matter is
\begin{equation}\label{mass_conservation}
\dot{\rho}_m+3H\left(\rho_m+p_m\right)=0
\end{equation}
and that of radiation is
\begin{equation}\label{radiation_conservation}
\dot{\rho}_r+3H\left(\rho_r+p_r\right)=0
\end{equation}
where, $G_{cosmo}=\frac{\kappa^2}{16\pi \left(3\lambda-1\right)}$
with the conditions $\frac{\kappa^4 \mu^2
\Lambda}{8\left(3\lambda-1\right)}=1$ and
$G_{grav}=\frac{\kappa^2}{32 \pi}$.\\

Solving the DE conservation equation for variable modified
Chaplygin Gas, the expression for DE density for VMCG will be
\begin{equation}\label{vmcg_density}
\rho_{vmcg}=\rho_{vmcg0}
(1+z)^{3(1+n)}\left[A+(1-A)(1+z)^{\gamma}\right]^{\frac{1}{\alpha+1}}.
\end{equation}
where
$\rho_{vmcg0}=\left[-\frac{3\left(n+1\right)\beta_{0}}{\gamma}+C\right]^\frac{1}{\alpha+1}$,~~
$A=\frac{C}{\rho_{vmcg0}^{\alpha+1}}$, where $C$ is the
integration constant.

%%%%%%%%%%%%%%%%%%%%%%%%%%%%%%%%%%%%%%%%%%%%%%%%%%%%%%%%%%%%%%%%%%%%%%%%%%%%%%%%%%%%%%%%%%%%%%%%%%%%%%%%%%%%%%%%%%%%%%%%%%%%%%%%%%%%%
\section{Observational Constraints on EOS Parameters}\label{data}
%%%%%%%%%%%%%%%%%%%%%%%%%%%%%%%%%%%%%%%%%%%%%%%%%%%%%%%%%%%%%%%%%%%%%%%%%%%%%%%%%%%%%%%%%%%%%%%%%%%%%%%%%%%%%%%%%%%%%%%%%%%%%%%%%%%%%%%%
Using Eqs. (10)-(11), the Friedmann's equation can be rewritten
as:
\begin{equation}\label{HLFriedmann1}
H^{2}=\frac{8\pi
G}{3}(\rho_{vmcg}+\rho_{dm}+\rho_{r})+\left(\frac{K^{2}}{2\Lambda
a^{4}}+\frac{\Lambda}{2}\right)-\frac{K}{a^{2}}
\end{equation}

\begin{equation}\label{HLFriedmann2}
\dot{H}+\frac{3H^2}{2}=-4\pi G \left(p_{vmcg}
+\frac{1}{3}\rho_{r}\right)-\left(\frac{K^{2}}{4\Lambda
a^{4}}-\frac{3\Lambda}{4}\right)-\frac{K}{2a^{2}}
\end{equation}
We define the following dimensionless density parameters:

(i) for matter component:

\begin{equation}
\Omega_{i0}=\frac{8\pi G}{3H^{2}}\rho_{i0}
\end{equation}

(ii) for curvature:

\begin{equation}
\Omega_{K0}=-\frac{K}{H_{0}^{2}}
\end{equation}

(iii) for cosmological constant:

\begin{equation}
\Omega_{0}=\frac{\Lambda}{2H_{0}^{2}}
\end{equation}

Another dimensionless parameter for expansion rate is defined as:
\begin{equation}
E(z)=\frac{H(z)}{H_{0}}
\end{equation}
Using the above parameters, the Fridmann equation can be rewritten
as:
\begin{equation}
E^{2}(z)=\Omega_{vmcg0} F(z)+\Omega_{dm0}(1+z)^{3}
+\Omega_{r0}(1+z)^{4}+\frac{\Omega_{K0}^{2}}{4\Omega_{0}}(1+z)^{4}
+\Omega_{0}+\Omega_{K0}(1+z)^{2}
\end{equation}
where
\begin{equation}
F(z)=\rho_{vmcg0}(1+z)^{3(1+n)}\left[A+(1-A)(1+z)^{\gamma}\right]^{\frac{1}{\alpha+1}}
\end{equation}
Here $\Omega_{r0}$ is for radiation and $\Omega_{dm0}$ is for dark
matter. At the present epoch $E(z=0)=1$, which leads to
\begin{equation}
\Omega_{vmcg0}+\Omega_{dm0}+\Omega_{r0}+\frac{\Omega_{K0}^{2}}{\Omega_{0}}+\Omega_{0}+\Omega_{K0}=1
\end{equation}
%%%%%%%%%%%%%%%%%%%%%%%%%%%%%%%%%%%%%%%%%%%%%%%%%%%%%%%%%%%%%%%%%%%%%%%%%%%%%%%%%%%%%%%%%%%%%%%%%%%%%%%%%%%%%%%%%%%%%%%%%%%%%%%%%%%%%
\section{Observational Data Analysis In HL Universe}\label{data}
%%%%%%%%%%%%%%%%%%%%%%%%%%%%%%%%%%%%%%%%%%%%%%%%%%%%%%%%%%%%%%%%%%%%%%%%%%%%%%%%%%%%%%%%%%%%%%%%%%%%%%%%%%%%%%%%%%%%%%%%%%%%%%%%%%%%%%%%
In this section, we perform a detailed observational data analysis
\cite{Thakur1, Paul3, Paul4, Chakraborty1, Ghose1} using Stern
data. We also study the model under Stern+BAO and Stern+BAO+CMB
joint analysis. The mechanism that we will use in the present work
is the $\chi^{2}$ minimum test from theoretical Hubble parameter
with the observed data set and find the best fit values of unknown
parameters for different confidence levels (66\%, 90\%, 99\%). In
the table given below, we present the 3 column Stern data.

\[
\begin{tabular}{|c|c|c|}
\hline
  ~~~~~~$z$ ~~~~& ~~~~$H(z)$ ~~~~~& ~~~~$\sigma(z)$~~~~\\
  \hline
  0 & 73 & $\pm$ 8 \\
  0.1 & 69 & $\pm$ 12 \\
  0.17 & 83 & $\pm$ 8 \\
  0.27 & 77 & $\pm$ 14 \\
  0.4 & 95 & $\pm$ 17.4\\
  0.48& 90 & $\pm$ 60 \\
  0.88 & 97 & $\pm$ 40.4 \\
  0.9 & 117 & $\pm$ 23 \\
  1.3 & 168 & $\pm$ 17.4\\
  1.43 & 177 & $\pm$ 18.2 \\
  1.53 & 140 & $\pm$ 14\\
  1.75 & 202 & $\pm$ 40.4 \\ \hline
\end{tabular}
\]
{\bf Table 1:} The Hubble parameter $H(z)$ and the standard error
$\sigma(z)$ for different values of redshift $z$.\\

\begin{figure}
\includegraphics[height=2in]{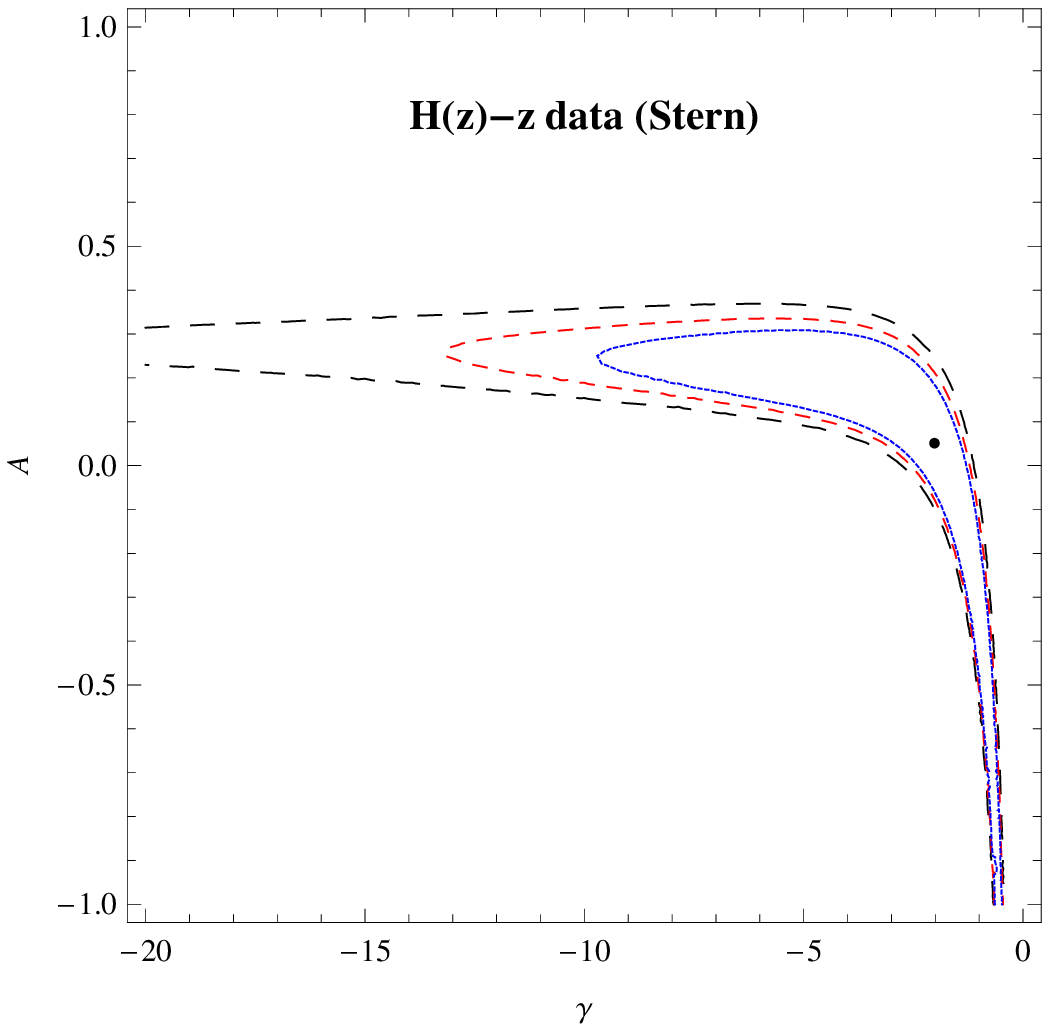}~~
\includegraphics[height=2in]{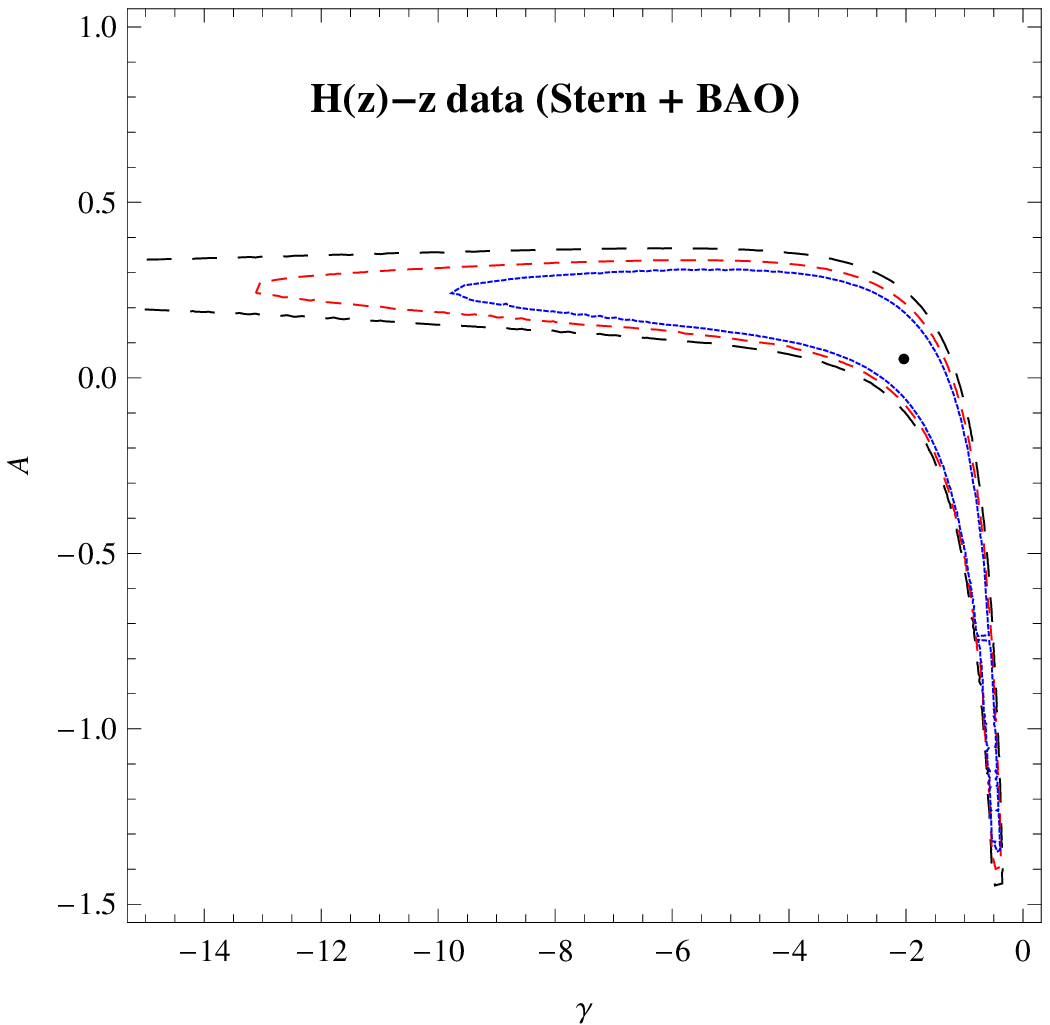}~~
\includegraphics[height=2in]{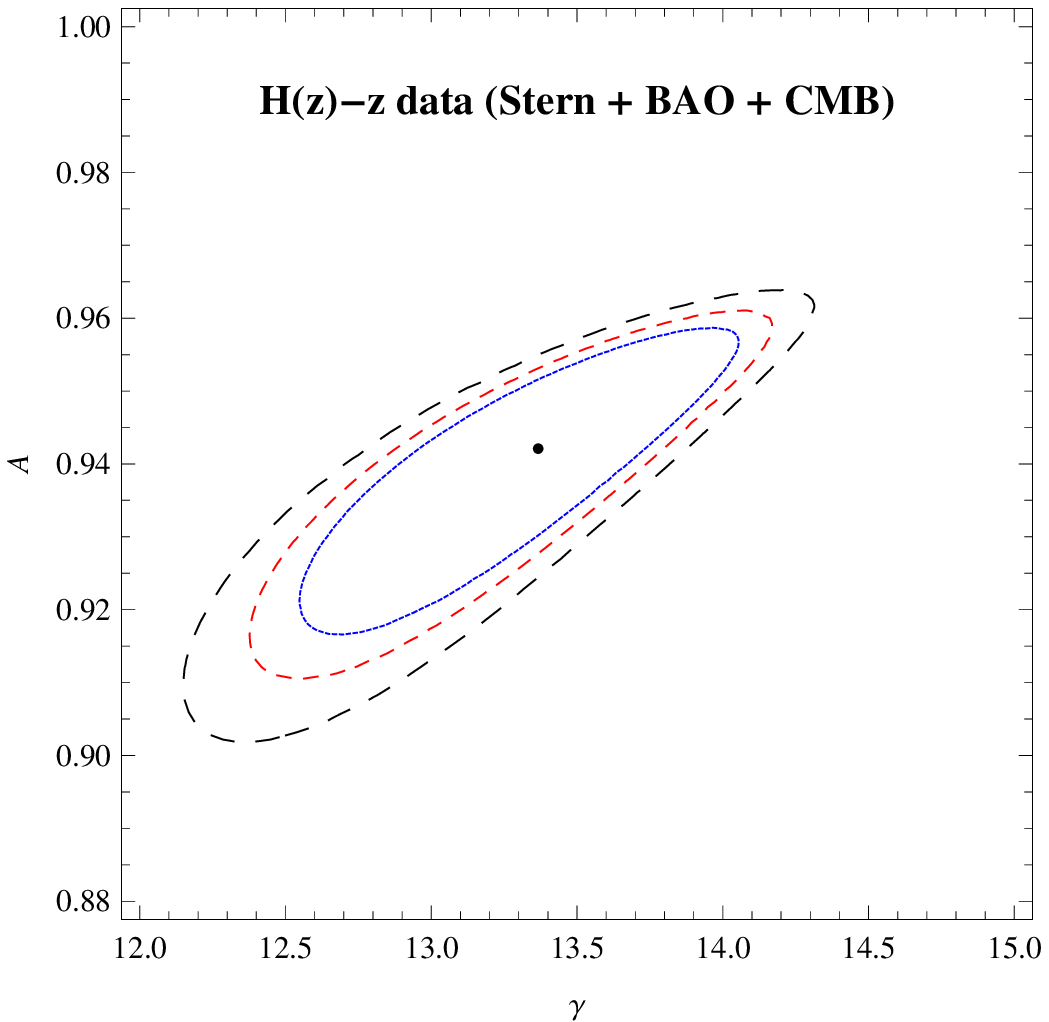}\\
\vspace{2mm}
~~Fig.1~~~~~~~~~~~~~~~~~~~~~~~~~~~~~~~~~~~~~~~~~Fig.2~~~~~~~~~~~~~~~~~~~~~~~~~~~~~~~~~~~~~~~~Fig.3\\
\vspace{1mm} The contours are drawn for 66\% (solid, blue), 90\%
(dashed, red) and 99\% (dashed, black) confidence levels.\\

Fig.1 shows the variations of $A$ against $\gamma$ for
$\alpha=0.001,n=0.1, \Omega_{i0}=0.01, \Omega_{r0}=0.02,
\Omega_{K0}=0.01, \Omega_{0}=0.01, \Omega_{dm0}=.03 $ with
respectively for the
$H(z)$-$z$ joint analysis.\\

Fig.2 shows the variations of $A$ against $\gamma$ for
$\alpha=0.001,n=0.1, \Omega_{i0}=0.01, \Omega_{r0}=0.02,
\Omega_{K0}=0.01, \Omega_{0}=0.01, \Omega_{dm0}=.03 $ with
respectively for the $H(z)$-$z$+BAO joint analysis.\\

Fig.3 shows the variations of $A$ against $\gamma$ for
$\alpha=0.001,n=0.1, \Omega_{i0}=0.01, \Omega_{r0}=0.02,
\Omega_{K0}=0.01, \Omega_{0}=0.01, \Omega_{dm0}=.03 $ with
respectively for the $H(z)$-$z$+BAO+CMB joint analysis.
 \vspace{1mm}

\end{figure}

\subsection{Stern ($H(z)$-$z$) Data Set}
In this sub-section our theoretical model of VMCG in HL gravity is
analyzed, using the observed values of Hubble parameter at
different redshifts (twelve data points) listed in observed Hubble
data by Stern et al \cite{Stern1}. The observed values of Hubble
parameter $H(z)$ and the standard error $\sigma(z)$ for different
values of redshift $z$ are given in Table 1. A statistical
hypothesis is proposed and its validity is tested at different
confidence levels. For this purpose we first form the $\chi^{2}$
statistics as a sum of standard normal distribution as follows:
\begin{equation}
{\chi}_{Stern}^{2}=\sum\frac{(H(z)-H_{obs}(z))^{2}}{\sigma^{2}(z)}
\end{equation}

\begin{equation}
L= \int e^{-\frac{1}{2}{\chi}_{Stern}^{2}}P(H_{0})dH_{0}
\end{equation}

where $H(z)$ and $H_{obs}(z)$ are theoretical and observational
values of Hubble parameter at different redshifts respectively.
Here, $H_{obs}$ is a nuisance parameter and can be safely
marginalized. $H_{0}$ is the present value of Hubble parameter and
its value is fixed at $H_{0}$ = 72 $\pm$ 8 Kms$^{-1}$ Mpc$^{-1}$.
From the DE model we see that the two most important parameters
are $A$ and $B(a)$. Here we shall determine best fit value of the
parameters ($A$ vs $\gamma$) by minimizing the above distribution
${\chi}_{Stern}^{2}$ and fixing the other unknown parameters with
the help of Stern data. According to our analysis the best fit
values of $A$ vs $\gamma$ is presented in Table 2. We now plot the
graph for different confidence levels. Our best fit analysis with
Stern observational data support the theoretical range of the
parameters. The 66\% (solid, blue), 90\% (dashed, red) and 99\%
(dashed, black) contours are plotted in fig.1 for
$\alpha=0.001,n=0.1, \Omega_{i0}=0.01, \Omega_{r0}=0.02,
\Omega_{K0}=0.01, \Omega_{0}=0.01, \Omega_{dm0}=.03 $.

\subsection{Stern $+$ BAO Data Sets}
Here we resort to a joint analysis in the sense that the BAO peaks
are incorporated in the stern data. The Baryon Acoustic
Oscillation (BAO) peak parameter value was proposed by
\cite{Eisenstein} and we shall use their approach. The pioneer as
far as the detection of BAO signal is concerned, is considered to
be the Sloan Digital Sky Survey (SDSS) survey. The survey directly
detected the BAO signals at a scale $\sim$ 100 MPc. The analysis
that is followed is actually the combination of angular diameter
distance and Hubble parameter at that redshift. This analysis is
independent of the measurement of $H_{0}$ and does not contain any
particular dark energy. Here we examine the parameters $A$ vs
$\gamma$ for VMCG model from the measurements of the BAO peak for
low redshift (with range $0<z<0.35$) using standard $\chi^{2}$
analysis. The error corresponds to the standard deviation, where
the distribution considered is Gaussian. We know that the
Low-redshift distance measurements are very lightly dependent on
different cosmological parameters, the EoS of dark energy and have
the ability to measure the Hubble constant $H_{0}$ directly. The
BAO peak parameter is defined by
\begin{equation}
{\cal
A}=\frac{\sqrt{\Omega_{m}}}{E(z_{1})^{1/3}}\left(\frac{1}{z_{1}}~\int_{0}^{z_{1}}
\frac{dz}{E(z)}\right)^{2/3}
\end{equation}
Here $E(z)=H(z)/H_{0}$ is the normalized Hubble parameter, the
redshift $z_{1}=0.35$ is the typical redshift of the SDSS sample
and the integration term is the dimensionless comoving distance
for the redshift $z_{1}$. The value of the parameter ${\cal A}$
for the flat model of the universe is given by ${\cal A}=0.469\pm
0.017$ using SDSS data \cite{Eisenstein} from luminous red
galaxies survey. Now the $\chi^{2}$ function for the BAO
measurement is given as,

\begin{equation}
\chi^{2}_{BAO}=\frac{({\cal A}-0.469)^{2}}{(0.017)^{2}}
\end{equation}

The total joint data analysis (Stern+BAO) for the $\chi^{2}$
function may be defined by

\begin{equation}
\chi^{2}_{total}=\chi^{2}_{Stern}+\chi^{2}_{BAO}
\end{equation}

According to our analysis the best fit values of $A$ vs $\gamma$
for the joint scheme is presented in Table 2. Finally we generate
the closed contours of $A$ vs $\gamma$ for the 66\% (solid, blue),
90\% (dashed, red) and 99\% (dashed, black) confidence limits
depicted in fig.2 for $\alpha=0.001,n=0.1, \Omega_{i0}=0.01,
\Omega_{r0}=0.02, \Omega_{K0}=0.01, \Omega_{0}=0.01, \Omega_{dm0}=.03 $.\\

\subsection{Stern $+$ BAO $+$ CMB Data Sets}
The angular scale of the first acoustic peak is measured through
angular scale of the sound horizon at the surface of last
scattering. This is one of the most interesting geometrical probe
of dark energy. The information is encoded in the CMB (Cosmic
Microwave Background) power spectrum. The definition of the CMB
shift parameter is given in \cite{Bond1, Efstathiou1, Nessaeris1}.
It is not sensitive with respect to perturbations but are suitable
to constrain model parameter. This is the property that we will
use in this analysis. The CMB power spectrum first peak is the
shift parameter which is given by
\begin{equation}
{\cal R}=\sqrt{\Omega_{m}} \int_{0}^{z_{2}} \frac{dz}{E(z)}
\end{equation}
where $z_{2}$ is the value of redshift at the last scattering
surface. From WMAP7 data of the work of Komatsu et al
\cite{Komatsu1} the value of the parameter was obtained as ${\cal
R}=1.726\pm 0.018$ at the redshift $z=1091.3$. The $\chi^{2}$
function for the CMB measurement can be written as
\begin{equation}
\chi^{2}_{CMB}=\frac{({\cal R}-1.726)^{2}}{(0.018)^{2}}
\end{equation}

Now when we consider three cosmological tests together, the total
joint data analysis (Stern+BAO+CMB) for the $\chi^{2}$ function is
defined by

\begin{equation}
\chi^{2}_{TOTAL}=\chi^{2}_{Stern}+\chi^{2}_{BAO}+\chi^{2}_{CMB}
\end{equation}
Now the best fit values of ($A$,$\gamma$) for joint analysis of
BAO and CMB with Stern observational data support the theoretical
range of the parameters are given in Table 2. The 66\% (solid,
blue), 90\% (dashed, red) and 99\% (dashed, black) contours are
plotted in fig.3 for $\alpha=0.001,n=0.1, \Omega_{i0}=0.01,
\Omega_{r0}=0.02, \Omega_{K0}=0.01, \Omega_{0}=0.01,
\Omega_{dm0}=.03 $.

\[
\begin{tabular}{|c|c|c|c|}
\hline
  ~~~~~~$Data$ ~~~~~&~~~~~~~$A$ ~~~~~~~~&~~~~~~~~$\gamma$~~~~~&~~~~~$\chi^{2}_{min}$~~~~~~\\
  \hline
  ~~~~~~$Stern$ ~~~~~&~~~~~$0.0512871$ ~~~~~~&~~~$-2.02163$~~~~~&~~~~~~~~$7.17375$~~~~~~~~~~~\\
  \hline
  ~~~~~$Stern+BAO$ ~~~&~~~~~$0.0536315$ ~~~~~~&~~~$-2.03587$~~~~~&~~~~~$768.149$~~~~~~\\
  \hline
  ~~~~$Stern+BAO+CMB$~~&~~~~~~$0.942108$ ~~~~~~&~~~~$13.3671$~~~~~&~~~~~$9554.2$~~~~~~\\
   \hline
\end{tabular}
\]
{\bf Table 2:} The best fit values of $A$, $\gamma$ and the
minimum values of $\chi^{2}$.

\subsection{Redshift-Magnitude Observations from Supernovae Type Ia}

The main evidence for the existence of dark energy was provided by
Supernova Type Ia experiments. The existence of dark energy is
directly related to the redshift of the universe. Therefore, since
1995 two teams of High redshift Supernova Search and the Supernova
Cosmology Project have been working extensively, and in their
effort they have discovered several type Ia supernovae at the high
redshifts \cite{Perlmutter, Riess, Riess1, Perlmutter1}. The
observations directly measure the distance modulus of a Supernovae
and its redshift $z$ \cite{Riess, Kowalaski1}. Here we will
consider the recent observational data, including SNe Ia which
consists of 557 data points and belongs to the Union2 sample
\cite{Amanullah1}. From the observations, the luminosity distance
$d_{L}(z)$ will determine the dark energy density which is defined
by

\begin{equation}
d_{L}(z)=(1+z)H_{0}\int_{0}^{z}\frac{dz'}{H(z')}
\end{equation}

The distance modulus (distance between absolute and apparent
luminosity of a distance object) for Supernovae is given by

\begin{equation}
\mu(z)=5\log_{10} \left[\frac{d_{L}(z)/H_{0}}{1~MPc}\right]+25
\end{equation}
The best fit of distance modulus as a function $\mu(z)$ of
redshift $z$ for our theoretical model and the Supernova Type Ia
Union2 sample are drawn in fig.4 for our best fit values of the
parameters. From the curve, we see that the theoretical VMCG model
in HL gravity is in agreement with the union2 sample data.

\begin{figure}
~~~~~~~~~~~~~~~~~~~~~~~~\includegraphics[height=2in]{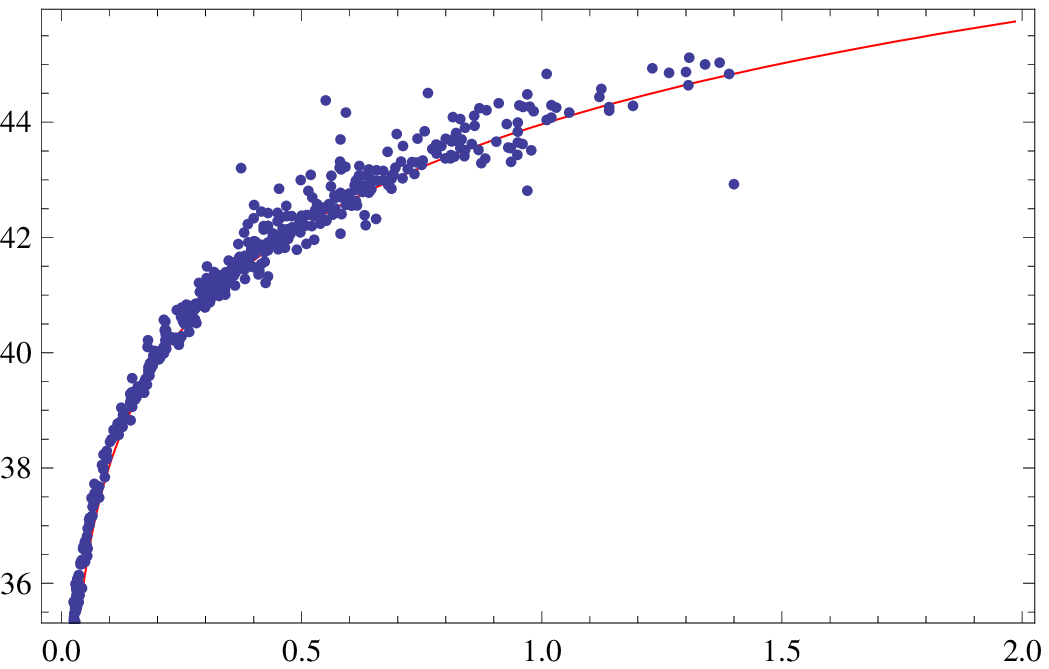}~~~~\\
\vspace{1mm} ~~~~~~~~~~~~~~~~~~~~~~~~~~~~~~~Fig.4: $\mu(z)$ vs $z$
for $n=0.1, \alpha=0.01$ \vspace{1mm}

\end{figure}

\section{\normalsize\bf{Discussions}}
In this work, we have considered the FRW universe in HL gravity
model filled with a combination of dark matter and dark energy in
the form of variable modified Chaplygin gas (VMCG). We present the
Hubble parameter in terms of the observable parameters
$\Omega_{i0}$, $\Omega_{K0}$, $\Omega_{0}$, $\Omega_{dm0}$,
$\Omega_{vmcg0}$ and $H_{0}$ with the redshift $z$ and the other
parameters like $A$, $n$, $\gamma$ and $\alpha$. For these
parameters we have chosen the specific numerical values consistent
with observations. From Stern data set (12 points), we have
obtained the bounds of the arbitrary parameters by minimizing the
$\chi^{2}$ test. Next due to joint analysis of BAO and CMB
observations, we have also obtained the best fit values and the
bounds of the parameters ($A,\gamma$). We have plotted the
statistical confidence contour of ($A,\gamma$) for different
confidence levels i.e., 66\%(dotted, blue), 90\%(dashed, red) and
99\%(dashed, black) confidence levels by fixing observable
parameters $\Omega_{dm0}$, $\Omega_{r0}$, $\Omega_{K0}$,
$\Omega_{0}$ and $H_{0}$ and some other parameters like, $n$ and
$\alpha$, etc. for Stern, Stern+BAO and Stern+BAO+CMB data
analysis.

From the Stern data,the best-fit values and bounds of the
parameters ($A,\gamma$) are obtained. The output values are shown
in Table 1 and the figure 1 shows statistical confidence contour
for 66\%, 90\% and 99\% confidence levels. Next due to joint
analysis with Stern + BAO data, we have also obtained the best-fit
values and bounds of the parameters ($A,\gamma$). The results are
displayed in the second row of Table 1 and in figure 2 we have
plotted the statistical confidence contour for 66\%, 90\% and 99\%
confidence levels. After that, due to joint analysis with
Stern+BAO+CMB data, the best-fit values and bounds of the
parameters ($A,\gamma$) are found and the results are shown in
Table 1. The figure 3 shows statistical confidence contour for
66\%, 90\% and 99\% confidence levels. For each case, we compare
the model parameters through their values and by the statistical
contours. From the comparative study, one can get an idea about
the convergence of theoretical values of the parameters with their
values obtained from the observational data set and how it is
altered for different chosen set of other parametric values.

Finally the distance modulus $\mu(z)$ against redshift $z$ has
been drawn in figure 4 for the theoretical model of the VMCG in HL
gravity for the best fit values of the parameters and the observed
SNe Ia Union2 data sample. So the observational data sets are
perfectly consistent with our predicted theoretical VMCG model in
HL gravity.

The present study discover the constraint of allowed composition
of matter-energy by constraining the range of the values of the
parameters for a physically viable VMCG in HL gravity model. In a
nut-shell, the conclusion of this discussion suggests that even
though the quantum aspect of gravity have small effect on the
observational constraint, but the cosmological observation can put
upper bounds on the magnitude of the correction coming from
quantum gravity that may be closer to the theoretical expectation
than what one would expect.\\

\section*{Acknowledgments}

The authors are thankful to IUCAA, Pune, India for warm
hospitality where a part of the work was carried out. The Authors
acknowledge the anonymous referees for enlightening comments that
helped to improve the quality of the manuscript.\\

\end{document}